# Demonstration of Superconductor Shift Registers with Energy Dissipation Below Landauer's Thermodynamic Limit

Sergey K. Tolpygo, *Senior Member, IEEE*, Evan B. Golden, and Vasili K. Semenov

*Abstract*—We study energy dissipation and propagation of information encoded by Josephson vortices in two types of circular shift registers: a) a uniform register composed of sections of discrete Josephson transmission lines (JTL) forming a closed loop with a flux pump allowing to change the number of moving fluxons; b) a nonuniform register composed of sections of the regular JTL and sections of JTLs utilizing nSQUIDs – dc-SQUIDs with negative inductance between their arms – instead of single Josephson junctions. nSQUIDs are parametric devices with a flexible double-well potential that were proposed as components for reversible computing. For the uniform register, we demonstrate the energy dissipation per bit-shift operation below the Landauer's thermodynamic limit $E_T = k_B T \ln 2$ up to propagation delays of ∼ 0.7 ns, corresponding to the circular information motion with frequencies up to ∼ 1.4 GHz. This does not contradict Landauer's minimum energy requirement for computations since information is not destroyed. For the nonuniform register, we find the minimum energy dissipation per bit shift of about $16E_T$ and attribute this to a nonuniform movement of vortices and energy barriers between the regular JTL and nSQUID sections. Differences of Josephson vortex propagation in both types of circular registers are discussed based on the measured current-voltage characteristics, extracted effective resistance and the terminal speed of Josephson vortices, and their dependences on the number of moving vortices. nSQUID inductance connecting JJs to the ground leads to an unusual type of lossless discrete transmission line with frequency-dependent impedance and propagation speed, both different from the regular JTLs.

*Index Terms*—Josephson junctions, Josephson transmission lines, Josephson vortices, long Josephson junctions, superconducting transmission lines, superconducting electronics, SFQ circuits

## I. INTRODUCTION

SUPERCONDUCTOR electronics appear to be the most energy efficient due to the possibility to eliminate all sources of energy dissipation except those related to the

This material is based upon work supported by the Under Secretary of War for Research and Engineering under Air Force Contract No. FA8702-15-D-0001 or FA8702-25-D-B002. *(Corresponding author: Sergey K. Tolpygo).*

S.K. Tolpygo and E.B. Golden are with MIT Lincoln Laboratory, Lexington, MA 02421, USA (e-mail: sergey.tolpygo@ll.mit.edu). E. B. Golden is also with the Department of Electrical Engineering and Computer Science, Massachusetts Institute of Technology, Cambridge, MA 02139 USA (e-mail: ev27470@mit.edu).

V.K. Semenov is with the Department of Physics and Astronomy, Stony Brook University, Stony Brook, NY 11794, USA (e-mail: vasili.semenov@stonybrook.edu).

Color versions of one or more of the figures in this article are available online at http://ieeexplore.ieee.org

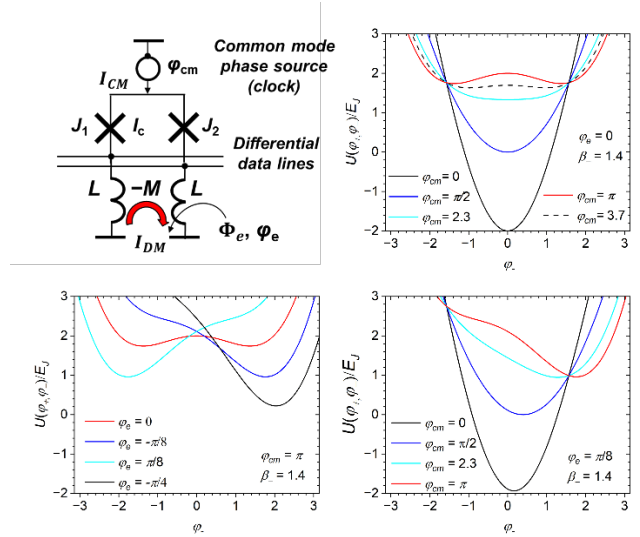

**Fig. 1.** Circuit diagram of an nSQUID consisting of two identical Josephson junctions, $J_1$ and $J_1$, with critical current $I_c$, inductance of the arms $L$ and mutual inductance between them $-M$. The nSQUID potential energy is controlled by a common mode phase source, $\varphi_{cm}$, and an external flux $\Phi_e$ which induces a differential mode current $I_{DM}$ and can be applied using differential data lines. Evolution of the nSQUID potential energy (1) as a function of $\varphi_- = (\varphi_1 - \varphi_2)/2$ at different values of the common mode phase bias $\varphi_{cm}$ and differential mode phase bias $\varphi_e = 2\pi \Phi_e/\Phi_0$ are shown for the case $\varphi_+ = \varphi_{cm}$; see text.

dynamics of its operation. Using quasi-adiabatic switching [1], [2] between states in a double-well potential provides the lowest energy dissipation per bit, approaching the minimum energy required for computations – the so-called Landauer's thermodynamic limit $E_T = k_B T \ln 2$ [3]. Although many superconducting devices utilizing adiabatic transitions have been demonstrated [4], [5], [6], [7], [8], [9], [10], [11] bit energies of $E_T$ have not been achieved.

It is also well known that Landauer's minimum energy requirement applies only to computations destroying information and increasing entropy [3], [12] whereas transporting or processing information without destroying it, e.g., in a circular shift register, could theoretically be done without any energy consumption. Reversible computing with physically and logically reversible bits has long been proposed as an ultimate solution for energy efficiency, short only of quantum computing; see [12] for a review. A number of reversible computing solutions with superconducting circuits has been proposed [13], [14], [15], [16], [17], [18], [19], [20], [21] and references therein. Note that quantum circuits should be physically and logically reversible, by definition.

In classical circuits, the lowest energy dissipation so far has been demonstrated in circuits based on nSQUIDs – adiabatic parametric devices comprised of a symmetric dc-SQUID with a negative mutual inductance, $-M$, between the SQUID arms [14], [15], [16], [17], [18]; see Fig. 1. The nSQUID potential energy is given by [14], [18], [48]

$$U(\varphi_+, \varphi_-)/E_J = [\frac{(\varphi_+ - \varphi_{cm})^2}{\beta_+} + \frac{(\varphi_- - \varphi_e)^2}{\beta_-} - 2 \cos \varphi_+ \cos \varphi_-],$$
$$(1)$$

where $E_J = I_c \, \Phi_0/2\pi$ is the Josephson coupling energy, $\varphi_\pm = (\varphi_1 \pm \varphi_2)/2$, $\varphi_1$, $\varphi_2$ are the Josephson phase differences across junctions $J1$ and $J2$, $\beta_\pm = 2\pi I_c (L \pm M)/\Phi_0$ are the dimensionless inductances; $\varphi_e = 2\pi \Phi_e/\Phi_0$ is the phase bias created by an external flux $\Phi_e$ applied to the nSQUID loop, and $\varphi_{cm}$ is the phase bias created by the common mode voltage source (or current $I_{CM}$ and the bias line inductance). The device can be designed to have a small total inductance $L_+ = (L - |M|)/2$ for the common bias current and a much larger inductance $L_- = 2(L + |M|)$ for the differential current $I_{DM}$ circulating the nSQUID loop, allowing for very different dynamics of the differential and common modes. In the absence of flux bias $\Phi_e$, the nSQUID has a symmetric double-well potential at $\varphi_{cm} = \pi$, corresponding to the clockwise and counter clockwise differential mode currents $I_{DM}$ circulating the loop. These two states are used to encode logic data. At the appropriate inductance parameters, transitions between these states can be done adiabatically through a monostable state in a single-well potential by controlling the value of $\varphi_{cm}$ and choosing the logic state in the left or right well by adjusting the sign and value of $\varphi_e$; see Fig.1 and [14], [15], [19] for more details.

Dynamics and description of an nSQUID are simplified if $L_- \gg L_+$, $\beta_+ \ll 1$, and $\varphi_{cm}$ is created by a voltage source $V(t)$. In this case $\varphi_+ \approx \varphi_{cm} \approx \int V(t) \, dt/\Phi_0$ and the potential for $\varphi_-$ dynamics oscillates with time from the double-well to the single-well and back, providing a suitable basis for reversible information processing, as shown in Fig. 1. In practical circuits, the voltage source is replaced by a Josephson transmission line (JTL), or by a long Josephson junction (LJJ), with a propagating train of Josephson vortices. The simplest nSQUID circuits – shift registers – present a long JTLs with some of the regular JJs (or a group of the regular junctions) replaced by nSQUIDs. The data signals to the nSQUIDs are provided via differential lines connected to the nSQUID arms and creating the appropriate flux $\varphi_e$ selecting the left or right well in the double-well potential. During the Josephson vortex passage through the nSQUID, the phase $\varphi_{cm}$ increases by $2\pi$ and energy profile (1) evolves through the full cycle, e.g., from a single-well at $\varphi_{cm} = 0$ to a double-well at $\varphi_{cm} = \pi$, and back to the single-well at $\varphi_{cm} = 2\pi$. In this way, Josephson vortices propagating on the JTL serve as a native clock for the nSQUIDs [13], [14], [15], [16], [17], [18], [19].

The data shift operation is the simplest operation in computing and does not destroy information. Hence, shift registers could theoretically operate with bit energies below $E_T$. Demonstrating this is a mandatory prerequisite for demonstrating more complex reversible operations with energies below $E_T$ since the data shift operation is always an essential part of any complex operation.

The simplest circuit allowing for a straightforward measurement of the energy dissipation is a circular shift register or a Josephson transmission line. The power dissipation in the JTL is simply $P = I_B V_d$, where $I_B$ is the JTL's dc bias current and $V_d$ the time-averaged voltage across the JTL. From the Josephson relation, $V_d = n\Phi_0/\tau$, where $n$ is the number of fluxons moving in the JTL ring and $\tau = 1/f$ is the period of their revolution. If the ring has $N$ Josephson junctions, all of them switch $n$ times during one period. So, the energy dissipation per $2\pi$ phase change across a single JJ is simply

$$E_{sw} = I_B \, \Phi_0/N \qquad (2)$$

To be useful as a clock source for the reversible nSQUID-based and quantum circuits, $E_{sw}$ should be less than $E_T$. Equivalently, the circuits should be able to operate at bias currents below a certain threshold

$$I_B \leq I_{th} \equiv Nk_BT \ln 2/\Phi_0, \qquad (3)$$

corresponding to about 20 nA per JJ at 4.2 K.

In this work we continue developing the nSQUID circuitry using a much more advanced circuit fabrication process [22] than the HYPRES process [23] used before, allowing for a substantial miniaturization of the nSQUIDs and increase in the circuit density and complexity. In II.A we start with a ring JTL composed of the regular Josephson junctions (JJs) and inductors, which can be viewed as a long annular JJ composed of discrete junctions. Dynamics of Josephson vortices in these circuits is reviewed in III, and the experimental results are given in IV. In II.B and V we introduce the design and test data for the register composed of nSQUIDs. All studied circuits were fabricated in the SFQ5ee process at MIT Lincoln Laboratory (MIT LL) [22], using Josephson critical current density of the junctions $j_c = 1$ μA/μm².

## II. DESCRIPTION OF THE CIRCUITS

### A. Uniform Register: Josephson Ring with Discrete Junctions

The first circuit is a Josephson ring oscillator comprised of $N = 256$ small, nominally identical Josephson junctions in a stripline configuration between two superconducting ground planes as shown in Fig. 2. This circuit is a discrete "circular" Josephson transmission line (JTL) or a discrete annular junction [44]. Its electrical schematic, layout, and cross section are shown in Fig. 2 along with the microphotograph of the fabricated circuit. The two ends of the stripline are joined together by a superconducting Flux Pump (FP) visible as a small shiny box at the bottom of Fig. 2. The JJs with target critical current $J_c = 15$ μA are equally spaced at 50 μm center-to-center; the stripline width is $w = 4$ μm and the total length

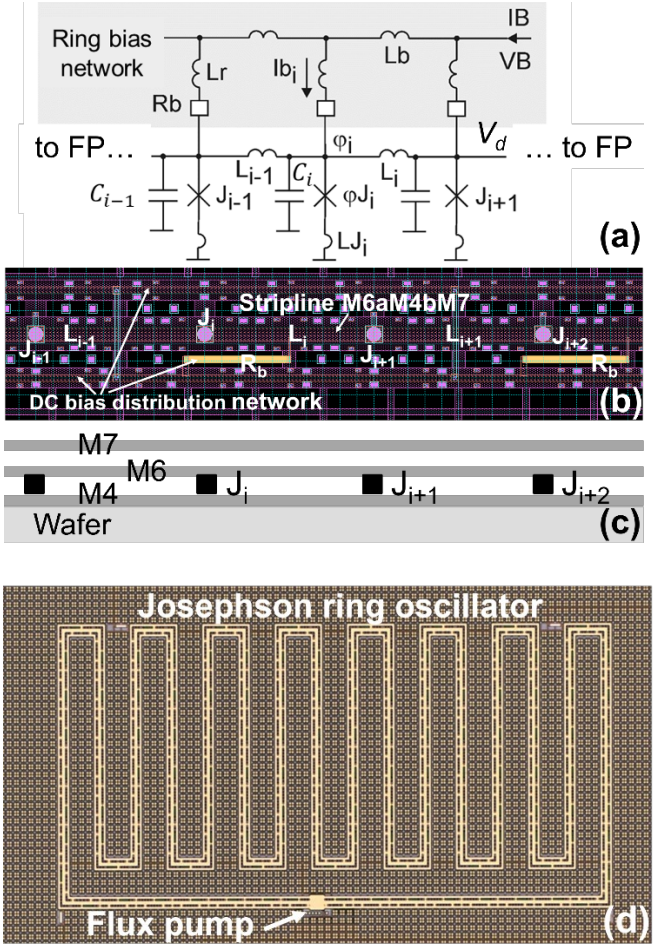

**Fig. 2.** Discrete circular Josephson transmission line (JTL) − the Josephson ring oscillator which can be viewed as a long circular Josephson junction comprised of $N$ individual small Josephson junctions connected by superconducting strips M6 between superconducting planes M4 and M7, striplines M6aM4bM7 in the SFQ5ee process [22], [23]. Panels from top to bottom show: a) a fraction of the circuit diagram explicitly showing three JJs, inductors and capacitors forming the transmission line, and the bias current network providing a uniform bias current distribution to individual JJs, using resistors $R_b$ and inductors $L_r$ and $L_b$; b) the actual layout of a few sections of the JTL; c) a schematic cross section of the M6aM4bM7 stripline, ground plane layers M4 and M7 are connected by superconducting vias (not shown); d) a microphotograph of the fabricated circuit referred hereafter Revcom4. All section of the stripline have the same inductance, $L_i = L_j \equiv L_{cell}$, and capacitance $C_i$ which is the sum of the stripline capacitance, $C_{cell}$ and the capacitance of $i$-th Josephson junction, $C_j$. M6 strip width, $w$ and the cell length, $p$ determine $L_{cell}$ and $C_{cell}$. $LJ_i$ in (a) is a small parasitic inductance associated with superconducting vias connecting junctions to the ground planes. The ends of the JTL are connected by a superconducting flux pump (FP).

$l = 12896\ \mu m$; see Table I.

The flux pump is a symmetrical two-junction SQUID that can operate in two modes; see Fig. 3. In the passive mode it emulates a tiny inductance closing JTL or nSQUID rings. In the active mode it is used to controllably inject flux vortices into the ring or extract them out. The device can be reprogrammed between the modes by changing a few control currents. In the passive mode, currents IM, IQ1 and IQ2 are equal to zero, and junctions J1 and J2 are in the superconductive state. The junctions are big enough to carry current IR flowing in the ring, which value depends on the number of vortices injected in the ring. For the flux pump operation, equal and opposite currents IQ1 = −IQ2 are applied to terminals IQ1 and IQ2 with respect to the common ground. Without changing the IR, this differential current sets the SQUID bias current $I_{bd}$ to the value corresponding to the intersection of the vertical dash red and the horizontal dash green lines in Fig. 3b. A new vortex is injected by a short current pulse applied to terminal IM, adding an external flux to the SQUID loop. The value of this current should be sufficient to cross the boundary of the "0" lob of the SQUID's critical current versus flux dependence but insufficient to escape from lob "1", so that the SQUID remains in the superconductive state. The injection of the vortex is completed when current IM is returned to zero. To withdraw the vortex, the procedure is repeated with the negative IM value.

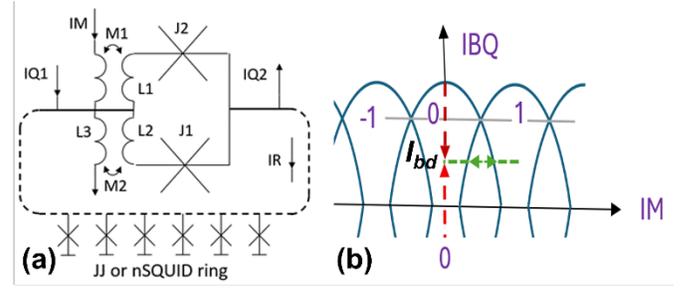

**Fig. 3.** The electrical schematic (a) and operation (b) of the flux pump (FP). The flux pump consists of a symmetrical dc-SQUID which can inject fluxons into the loop composed of the flux pump and the JTL or nSQUID ring. The flux injection is controlled by a differentially applied dc-SQUID bias current $I_{bd} = IQ1 - IQ2$ and the external flux created by magnetic bias current IM in a three-turn coil coupled to the dc-SQUID loop L1-L2; see text. FP design parameters: Ic1=Ic2=0.3 mA, L1=L2=3.1 pH, L3=53 pH, M1=M2=0.9 pH.

Designs of the JTL ring and the flux pump are similar to those used previously in [24], although we significantly miniaturized the flux pump with respect to its prior version. The circuit titled Revcom4 was fabricated in the SFQ5ee process with Josephson critical current density $j_c = 1\ \mu A/\mu m^2$.

### B. Nonuniform Register: Josephson Ring with nSQUIDs

The second circuit is also a ring similar to the first circuit but now containing also nSQUIDs. Specifically, the ring contains five sections of JTL composed of 26 nSQUIDs in each section. Each nSQUID section is connected to another nSQUID section by a JTL composed of 36 cells having regular dc-SQUIDs with small inductance instead of single JJs in II.A; see Fig. 4. In total, the ring has 192 dc-SQUID cells and 130 nSQUID cells, giving the total number of JJ pairs $N = 322$. The nSQUIDs differ from the regular dc-SQUIDs by the presence of additional inductors in the SQUID arms that have a negative mutual inductance $-M$, while the JJs and all other inductors are the same as well as the all stripline section inductors, $L_i = L_k = L_{cell}$. The flux pump joins two dc-SQUID-based JTL sections of the ring. The ring bias network and the flux pump are the same as in Fig. 2a and Fig. 3.

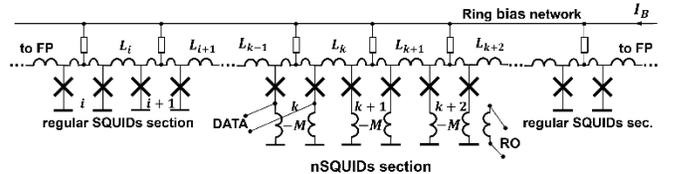

**Fig. 4.** A simplified electrical schematic of a Josephson ring with nSQUIDs, circuit Revcom5. The circuit consists of sections of regular Josephson transmission lines, using dc-SQUIDs with small inductance as shown on the left-hand side by devices $i$, $i + 1$, and sections of JTLs formed using nSQUIDs as shown in the middle by devices, $k, k + 1, k + 2$. The flux pump (FP) connects two ends of the ring's regular JTL sections.

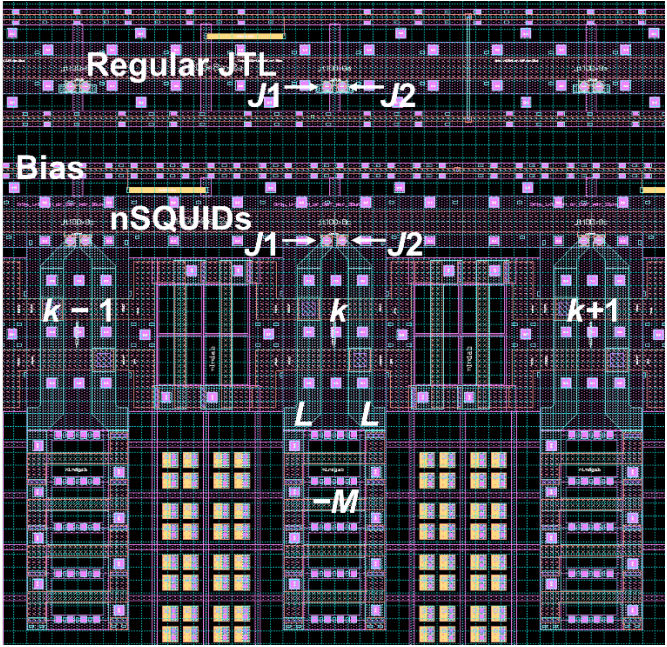

**Fig. 5.** Layout of a few sections of the shift register with nSQUIDs, circuit Revcom5a. It consists of sections of the regular JTL, similar to Fig. 2, and sections of the nSQUID-based JTLs. All sections use identical JJs $J1$ and $J2$ and the same placement pitch, $p$. Sections of the regular JTLs use dc-SQUIDs with negligibly small inductance, i.e., two parallel junctions, instead of one junction in section of Revcom 4, as shown on the left-hand side of Fig. 4 by devices $i$, $i+1$. Sections formed using nSQUIDs are shown in the middle by devices, $k-1, k, k+1$, having inductance $L$ in each arm and mutual inductance $-M$ between them.

Since inductance of the dc-SQUIDs in the ring is very small $\beta_L = 2\pi J_c L/\Phi_0 \ll 1$, each dc-SQUID behaves as two junctions in parallel with critical current $2J_c$. The circuit was designed for and fabricated in the process SFQ5ee process with Josephson critical current density $j_c = 1$ µA/µm². The size of the JJs was adjusted to get $2J_c = 16$ µA in one version of the circuit, Revcom5a, and $2J_c = 20$ µA in the second version, Revcom5b.

## III. Vortex Dynamics in Long Josephson Junctions

The equation for the phase difference between the banks of a long Josephson junction in the presence of a uniform bias current is the sine-Gordon equation [25], [26], [27]

$$\frac{\partial^2 \varphi}{\partial x^2} - \frac{1}{c_0^2}\frac{\partial^2 \varphi}{\partial t^2} - \frac{\sin \varphi}{\lambda_J^2} - \eta\frac{\partial \varphi}{\partial t} = i \qquad (4)$$

| Circuit | Junction area (µm²) | Junction capacitance, $C_J$ (pF) | Cell length, $p$ (µm) | Cell capacitance $C_{cell}$ (fF) | Inductances $L_{cell}$ and $L_+$ (pH) | $\beta_L$ | $\lambda_J$ (µm) | Total length (µm) and number of JJs |
|---|---|---|---|---|---|---|---|---|
| Revcom4 | 15 (single JJ) | 0.6ᵃ 0.89ᵇ | 60 | 50 | 3.83; n/a | 0.175 | 120 | 12896 $N$=256 |
| Revcom5, all JTL sections | 16 (two JJs in parallel) | 0.64ᵃ 0.949ᵇ | 100 | 228 | 4.01; n/a | 0.195 | 226 | 19300 $N$=192 |
| Revcom5 nSQUID sections | 16 (two JJs in parallel) | 0.64ᵃ 0.949ᵇ | 100 | 228 | 4.01; 6.6 | 0.516ᶜ | 226 139ᶜ | 13000, $N$=130 32300 (total ring, $N$=322) |

ᵃ Using junction specific capacitance of 40 fF/µm² from [45].
ᵇ Using junction specific capacitance of 59.3 fF/µm² that gives the best fit to the measured step voltage $V_0$ in the regular JTL ring.
ᶜ Assuming $L_+$ adds in series to $L_{cell}$.

Here $\lambda_J$ is Josephson penetration depth, $\eta$ is the viscosity coefficient, $i = J/J_c$ and $J_c$ are the normalized bias and the critical current of the long junction per unit length, and $c_0 = (LC)^{-1/2}$ the Swihart velocity [28] – the speed of electromagnetic waves in the junction, $L$ and $C$ inductance and capacitance per unit length, respectively. Solution of (1) without the friction term and the driving force (bias current) has a form of a moving soliton (vortex):

$$\varphi_0 = 4\tan^{-1} e^z, \qquad (5)$$

were $z$ and $\tau$ are the coordinate and time in the frame of the vortex moving with velocity $v$:

$$z = \frac{x - \beta t}{\gamma}, \tau = \frac{t - \beta x}{\gamma}, \beta = \frac{v}{c_0}, \text{and } \gamma = (1 - \beta^2)^{1/2} \qquad (6)$$

The bias current creates Lorentz force which accelerates the vortex until this force is balanced by the friction force proportional to the vortex speed. If viscosity coefficient is small, the stationary solution for the vortex is [28], [29], [30]

$$i = \frac{4\beta\eta}{\pi\gamma}. \qquad (7)$$

The average dc voltage across the junction is related to the rate of the phase change due to a moving vortex, $V = \Phi_0 v/l$, where $l$ the length of the junction. Converting (4) to dimensional units gives the long junction current-voltage characteristic (CVC) [29], [30], [31], [32], [33]

$$I = \frac{4v\eta l_c}{\pi c_0(1 - (v/c_0)^2)^{1/2}} = \frac{GV}{(1 - (V/V_0)^2)^{1/2}}, \qquad (8)$$

which looks like a giant current step at $V$ approaching $V_0$ [33]-[36], where $V_0 = \Phi_0 c_0/l$ is the voltage produced by a vortex moving with the speed of light in the junction (the Swihart speed), and $G$ is the effective conductance combining all energy losses of a moving vortex. Since the vortex energy grows indefinitely as its speed approaches $c_0$, the current required to maintain the vortex speed by transferring energy from the current source to the vortex also increases without limit.

Inverting (8) gives

$$V = \frac{V_0}{(1 + (I_0^2/I^2)^{1/2}}, \qquad (9)$$

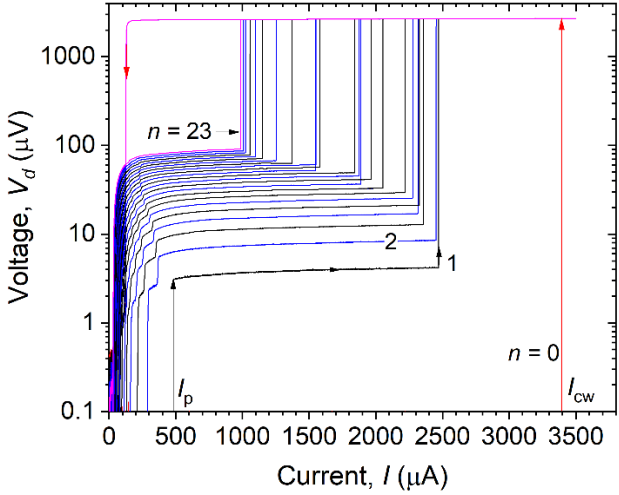

**Fig. 6.** Current-voltage characteristics (CVCs) of a Josephson ring, circuit Revcom4, consisting of $N = 256$ identical small unshunted Josephson junctions connected in parallel by inductors; see Fig. 2. The ring can be viewed as a circular Josephson transmission line with uniform distribution of the bias current, $I$ or as a circular shift register. In a vortex-free state, the ring switches from the zero-voltage state to the gap voltage $V_g$ =2.69 mV of the individual JJs at a switching current $I_{sw} \approx NI_c$, where $I_c$ is the critical current of the individual JJs comprising the ring. The ring re-traps back into the superconducting state at a much smaller current as shown by the left arrow. Adding Josephson vortices to the ring creates nearly horizontal, nearly equally spaced voltage steps from which the ring eventually switches to the gap voltage at currents smaller than the ring $I_{sw}$ as was first observed in [26], [34]; see text for details. The number of vortices in the ring, from 1 to 24, and the direction of the current sweeps are shown.

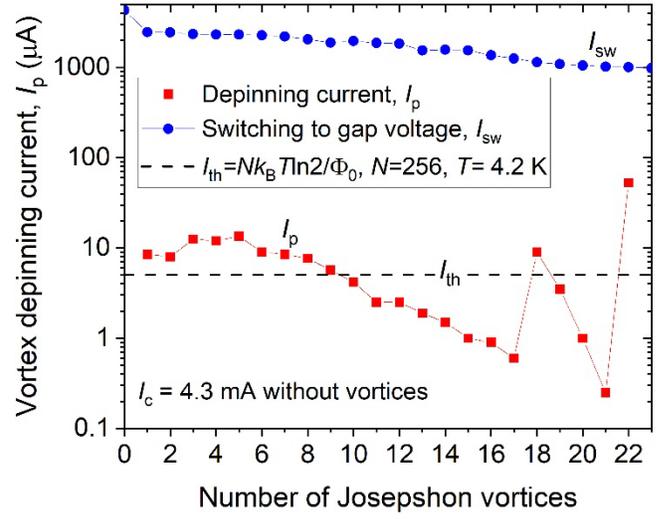

**Fig. 7.** Depinning critical current $I_p$ for Josephson vortices in the ring Revcom4b. The dash line shows the bias threshold current $I_{th} \equiv Nk_BT\ln 2/\Phi_0$ following from the Landauer's minimum energy dissipation $E_T$. Switching current to the gap voltage, $I_{sw}$ is also shown. In the range of bias currents $I_p \leq I_B \leq I_{cw}$, voltage on the JTL ring is caused by motion of Josephson vortices circulating the ring.

where $I_0 \equiv GV_0$. At small currents $I \ll I_0$, the CVC is linear $V = \frac{V_0}{I_0}I = R_0I$ with $R_0 \equiv G^{-1}$ the effective resistance of the ring to vortex motion. At large currents $I \gg I_0$, the CVC approaches the constant value $V = V_0$ (a voltage plateau on the CVC) corresponding to a relativistic vortex moving with the speed of light in the junctions.

When $n$ vortices are moving in the junction, the limiting (saturation) voltage is expected to be $V_n = nV_0$ and the effective small-current resistance is expected to be $R_{eff} = nR_0$ if the viscosity coefficient and damping do not depend on the number of moving vortices. The experimental CVCs in III were fit to extract $V_0$ and the effective resistance at small currents, $R_{eff}$. For the fitting, it turned out to be more convenient to convert (9) to an equivalent form

$$V = \frac{R_{eff}I}{(1+(I^2/I_0^2)^{1/2}} \qquad (10)$$

to avoid an apparent divergence of (9) at $I \to 0$.

The presence of nonuniformities of the ring, e.g., of the critical current of the individual junctions and/or inductors comprising the ring, modifies (4) and creates a vortex pinning potential [31], [32]. The corresponding pinning force prevents the vortex motion at currents $I \leq I_p$, where $I_p$ is the depinning current, often called the critical current because there is no voltage associated with vortex movement at $I \leq I_p$. If $I_p \ll I_c$, in the simplest approximation, the force accelerating the moving vortices is the difference between the Lorentz force

and the pinning force, which is proportional to $I - I_p$. Then, at currents $I \gg I_p$, (7) becomes

$$V = \frac{R_{eff}(I-I_p)}{(1+(I-I_p)^2/I_0^2)^{1/2}}, \qquad (11)$$

which we used to fit the experimental CVCs as shown in Figs. 8a,b. Thermal activation of vortices over the pinning barrier should cause some rounding of the CVCs at currents close to $I_p$ which will not be considered here.

For a ring composed of discrete JJs, (4) is replaced by a discrete sine-Gordon equation, see, e.g., [37], [38]

$$\frac{(2\varphi_i-\varphi_{i-1}-\varphi_{i+1})}{\beta_L} + \beta_c\frac{\partial^2\varphi_i}{\partial\tau^2} + \frac{\partial\varphi_i}{\partial\tau} + \sin\varphi_i = i_i, \quad (12)$$

$$i = 1,\dots,N$$

where $\varphi_i$ is the phase difference across the $i$th junction, $i_i = I_i/I_c$ is the normalized bias current of the $i$th junction, $I_c$ the junction critical current, $\beta_L = 2\pi L_{cell}I_c/\Phi_0$ and $\beta_c = 2\pi I_c R^2(C_J + C_{cell})/\Phi_0$ are Stewart-McCumber inductance and damping parameters, respectively. Here $R$ and $C_J$ are the junction shunting resistance and junction capacitance, respectively; $C_{cell}$ and $L_{cell}$ are capacitance and inductance, with respect to the ground, of the superconducting transmission line, length $p$, between the centers of the $i$th and $(i + 1)$th junctions. In (12), we use the normalized time $\tau = t/t_c$ with $t_c = \Phi_0/(2\pi I_cR)$. The speed of electromagnetic waves on the discrete ring is $v_0 = p\omega_0$, where $\omega_0^{-2} = L_{cell}(C_J + C_{cell})$. Josephson penetration depth is $\lambda_J = p/\beta_L^{1/2}$, which is equivalent to $\lambda_J$ in (4) since $L_{cell} = \ell p$ and $I_c = J_c p$, where $\ell$ and $J_c$ are inductance and critical current per unit cell length. For a uniformly distributed bias current $I_B$, $i_i = I_B/(NI_c)$. At low voltages, the junction internal shunting resistance, characterizing dissipation in (12), can be very different from the junction normal state resistance $R_N$. For the unshunted junctions which we use, $\beta_c \gg 1$ and, typically,

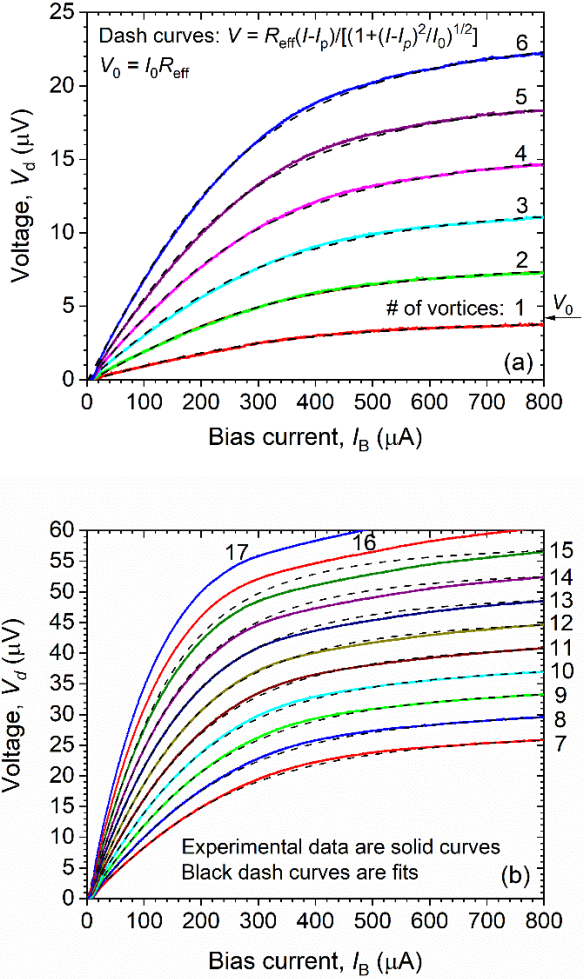

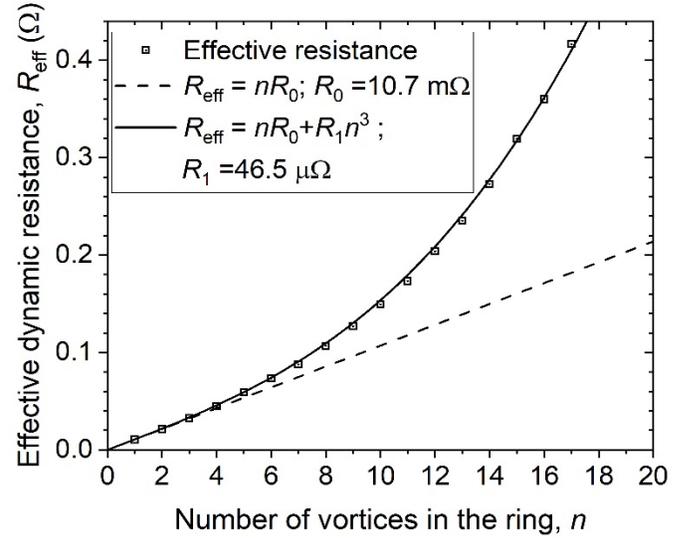

**Fig. 9.** Effective resistance $R_{eff}$ to a viscous vortex motion on the Josephson ring extracted from CVCs of RevCom4 circuit with different number of inserted vortices.

**Fig. 8.** CVCs of the Josephson ring Revcom4 with different number of inserted vortices: a) from 1 to 6; b) from 7 to 17. Experimental data are shown by colored curves. All black dash curves are fits to (11) with $I_0$ and $R_{eff}$ as fitting parameters. The product $I_0 R_{eff} = V_0$ corresponds to the voltage generated by a single relativistic vortex moving with the speed of light in the JTL.

$R \gg R_N$. In the small discreteness limit $p \to 0$, the first term in (12) becomes $-\Lambda_J^2 \partial^2 \varphi / \partial x^2$ and (12) reduces to (4).

## IV. EXPERIMENTAL RESULTS ON THE UNIFORM REGISTER

### A. Depinning Critical Current of the Ring JTL, Revcom4

The Revcom4 ring consisting of $N = 256$ individual junctions, each designed with the critical current $I_c = 15$ μA, is designed to have the total critical current $N I_c = 3.84$ mA. The measured switching current was $I_{sw} = 3.4$ mA, about 5% lower than the designed critical current value.

The depinning critical current of the ring with vortices is shown in Fig. 7. In an ideal long junction, a vortex can be moved and accelerated by an infinitely small bias current resulting in a voltage drop across the JJs. In the discrete JTL there is a finite barrier between the equivalent positions of a vortex in the JTL, which depends on the $\beta_L$ and vanishes at $\beta_L \to 0$. Ring nonuniformities and Abrikosov vortices trapped in or near the ring cells can cause pining of Josephson vortices. These all result in a finite depinning current, $I_p$ above which vortex motion starts and voltage on the ring appears. Hence, the $I_c/I_p$ ratio can characterize the ring uniformity. This ratio is about 430 for one vortex in the ring and grows by almost two orders of magnitude with increasing the number of vortices, indicating a very high uniformity of the fabricated ring.

Inspection of Fig. 7 shows that, at $n > 9$, the bias current required to move information in the circular shift register drops below the $I_{th}$, $I_p \leq 2\pi N k_B T \ln 2 / \Phi_0$, indicating that the energy dissipation per shift operation is below the Landaur's thermodynamic limit and that this circuit is worth investigating further.

Diminishing of $I_p$ with increasing $n$ is likely a result of repulsive interactions between the vortices, increasing their energy and making it easier to overcome the potential barrier by an ensemble of vortices than by a single vortex. Interesting commensurability effects in $I_p$ vs $n$ dependence have been observed in [24] at $n/N = 1/3$, 1/2, and 2/3. Unfortunately, we were not able to reach such large $n$ numbers because of the limitations in the flux pump procedure used.

### B. Current-Voltage Characteristics and Extracted Parameters

CVCs of the ring with different number of inserted vortices are shown in Fig. 8 along with the fits to (10)-(11). The effective resistance extracted from the fits is shown in Fig. 9. At a small number of vortices, the effective resistance is proportional to the number of moving vortices $R_{eff} = n R_0$ as expected for a completely linear medium, i.e., if one moving vortex does not change the properties of the medium, its viscosity coefficient, in which the next vortex moves. From Fig. 9, this holds up to approximately four vortices in the ring and then the effective resistance starts to quickly grow. The simplest polynomial fit gives $R_{eff} = n R_0 + R_1 n^3$ with $R_0 = 10.7$ mΩ and $R_1 = 46.5$ μΩ.

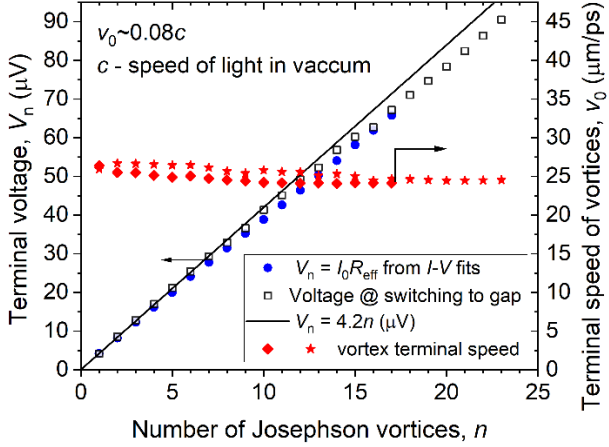

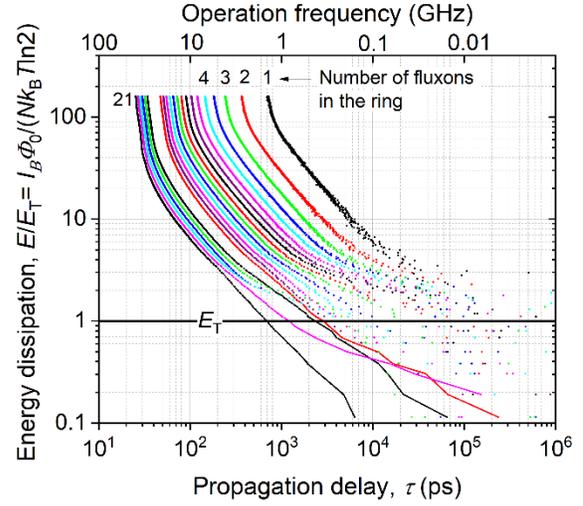

**Fig. 10.** Terminal voltage extracted from the CVCs fits as $V_n = I_0 R_{eff}$ (●) and as a voltage on the plateau right before the switching to the gap voltage state (□) as a function of the number of vortices moving in the ring. Also shown is the calculated terminal speed of vortices (right scale) using the values of the terminal voltage and the length of the JTL ring.

With bias current increasing, each CVCs in Fig. 8 saturates as voltage approaches the corresponding $V_n$ as a result of vortices approaching their ultimate speed $v_0$ on the JTL. To calculate $v_0$ we can use the voltage value right before the ring switches to the gap state, see Fig. 6, or use the terminal voltage extracted from fitting the CVCs to (10), (11) with $n$ vortices, $V_n = I_0 R_{eff}$. These two voltages are shown in Fig. 10 as a function of $n$. The linear fit of the data at small $n$ gives $V_n = 4.2n$ in microvolts, i.e., the value $V_0 = 4.2$ μV for a single vortex in the ring.

The terminal speed of vortices in the ring is
$$v_0 = l V_0 / \Phi_0. \qquad (13)$$
Using the ring length $l = 12896$ μm, we get $v_0 = 26.2$ μm ps$^{-1}$ or about 8.5% of the speed of light in vacuum. This speed can be compared with the speed of light in the passive transmission line, the Swihart speed, calculated from the ring cell design parameters in Table I,
$$c_0 = p / [L_{cell}(C_J + C_{cell})]^{1/2} \qquad (14)$$
and giving $c_0 = 31.4$ μm ps$^{-1}$; see Table II. Matching the value obtained from (13) requires the junction specific capacitance $C_s$ to be 59.3 fF/μm$^2$. The latter is a noticeably higher capacitance than the value $C_s \approx 40$ fF/μm$^2$ cited in numerous publications for the junctions with $j_c = 1$ μA/μm$^2$; see [45] and references therein. We will use the fitted $C_s$ value to model properties of the nonuniform registers with nSQUIDs. Note a gradual decrease in the voltage difference $V_{n+1} - V_n$ with respect to $V_0$ and the extracted speed of vortices $v_0$ with the number of moving vortices increasing. This decrease is qualitatively similar to changes in the positions of the Fiske mode resonances in one-dimensional discrete arrays observed in [38], [39], [40] and associated with bending of the dispersion relation of electromagnetic waves near the Brillouin zone edge [38]. We note in this regard that mutual inductance between the adjacent cells, which affects the dispersion relation [38], can be completely neglected in our discrete arrays due to the use of two connected superconducting ground planes, M4 and M7, below and about the JJs and inductors M6 in Fig. 2, which dramatically reduce all mutual inductances [41].

**Fig. 11.** Energy dissipated per Josephson junction in the circular register to move (shift) a bit of information (a Josephson vortex) around the register, $E = I_B \Phi_0 / N$, as a function of the propagation delay $\tau = n\Phi_0 / V_d\, V_n$ – the time it takes to travel around the ring – at different numbers of moving bits. For time delays larger than about 0.7 ns, corresponding to circulation frequency of about 1.4 GHz, the energy consumption is below Landauer's thermodynamic limit $E_T = k_B T \ln 2$. This does not contradict Landauer's minimum energy requirement since the information is not destroyed during the circular motion in the register.

### A. Energy Dissipation Versus Operation Frequency

The fact that energy dissipation in our circuit can be below $E_T$ has already been demonstrated by Fig. 7. However, criterion (3) and Fig. 7 do not tell us how fast information can be moved or processed without crossing $E_T$. This requires an energy-delay chart typically used to compare performance of various computational devices. The total energy dissipation in the ring during one revolution cycle, $E_{tot} = I_B V_d \tau$, where $\tau$ is the vortex (information) revolution period. The revolution period $\tau$ is the time delay between two successive deliveries of information to the same point on the ring. Since there are $N$ junctions and $n$ bit of information (vortices) being delivered, and $V_d = n\Phi_0 / \tau$, the energy dissipation per bit per JJ is simply $E = I_B \Phi_0 / N$. Fig. 11 shows this energy dissipation normalized to the Landauer's thermodynamic limit $E_T$ as a function of the time delay $\tau$ between two successive events of the information delivery around the ring. In the studied register, the minimum time delay, $\tau_{min}$ information can be moved without exceeding $E_T$ is about 0.7 ns, corresponding to the maximum circulation frequency of about 1.4 GHz and the maximum speed of information propagation $v_T = l/\tau_{min} \approx 18.4$ μm/ps that is about $0.7v_0$, 0.7 of the maximum propagation speed of Josephson vortices on the JTL ring.

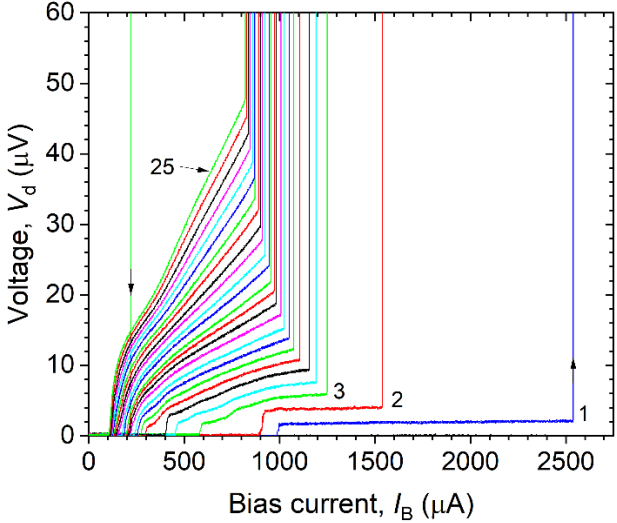

**Fig. 12.** CVCs of the Josephson ring Revcom5 composed of regular JTL sections and nSQUID-based sections, with different number of inserted vortices from 1 to 25, from right to left. A practically current-independent voltage plateaus occurs at a very small number of vortices from 1 to 3 after the bias current exceeds the depinning current.

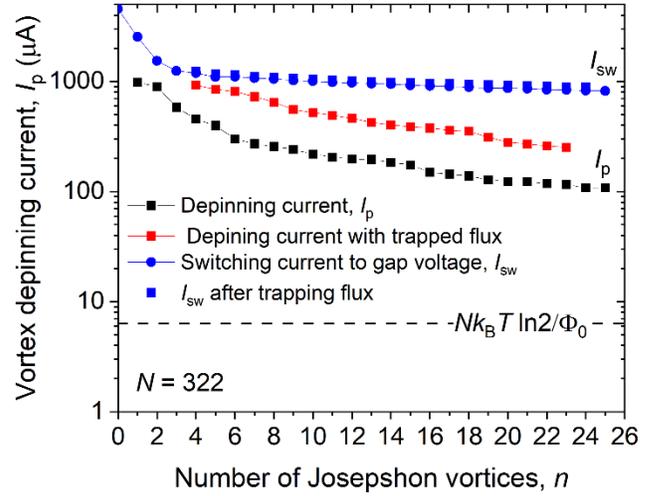

**Fig. 13.** Vortex depinning current, $I_p$ and switching current to the gap voltage, $I_{sw}$ for a nonuniform ring consisting of sections of the regular JTL and JTLs with nSQUIDs, circuit Revcom5, as a function of the number of inserted Josephson vortices. Magnetic flux trapped inside or near the circuit significantly affects the value of $I_p$ and the shape of the CVCs but usually do not change $I_{sw}$. The critical current of the ring without vortices (the switching current to the gap voltage) is 4.55 mA or 14 µA per SQUID that is close to the design value of 16 µA.

## V. EXPERIMENTAL RESULTS FOR THE NONUNIFORM REGISTER WITH nSQUIDS

### A. CVCs and Depinning Current of the Ring Composed of Regular JTLs and nSQUID-Based JTLs, Revcom5

CVCs of the register with nSQUIDs, described in II.B, are shown in Fig. 12 for the states with progressively increasing number of inserted Josephson vortices. Switching to the gap voltage (out of scale in Fig. 12) and re-trapping back into the vortex state are shown by the up and down arrows.

Comparison with Figs. 7 and 8 shows clear differences between the CVCs of the uniform JTL ring and the nonuniform ring consisting of sections of the regular JTLs and nSQUID JTLs:

a) the depinning current is much larger in the latter case;

b) voltage plateaus corresponding to Josephson vortices accelerating to the maximum speed exist only at a very small number of vortices: one, two and, perhaps, three.

c) at larger numbers of vortices, the CVCs resemble CVCs of a resistively and capacitively shunted junction above $I_p$ and have nearly linear $V_d(I_B)$ dependence before switching to the gap voltage;

d) a nearly stepwise increase in voltage at the $I_p$ threshold, contrary to a linear $V_d(I_B)$ increase with slope $R_{eff}$ in the regular JTL.

We also observed a much high sensitivity of the nSQUID register to flux trapping than in the regular JTL case, reflected in noticeable changes in the CVC shape and the value of the depinning current changing from cooldown to cooldown. The data in Figs. 12 and 13 show the lowest depinning current we were able to observed after multiple cooldowns in a triple-layer mu-metal shielded test probe with residual magnetic field of about 0.5 mG.

### B. Terminal Voltage and Propagation Speed in the Nonuniform Ring

The terminal voltage, defined as a voltage on the plateau corresponding to Josephson vortices moving with the terminal velocity in the ring, is clearly observed for just a few vortices in Fig. 12. To characterize the average propagation speed and compare with the data in Fig. 10, we used the voltage right before the ring switching to the gap-voltage state in place of the terminal voltage, $V_n$. As for the regular JTL ring, the voltage increases in steps proportionally to the number of inserted vortices; see Fig. 14. A liner fit $V_n = V_0 n$ gives $V_0 = 1.88$ µV. Using (13) and the total ring length $l = 32300$ µm, we get the average vortex propagation speed in the nonuniform register $< v_0 >$=29.4 µm/ps, a noticeably larger speed than in the Revcom4 ring in Fig. 10. However, this average speed is much lower that the Swihart speed calculated using the cell parameters in Table II, in the same manner as for Revcom4; see Sec. VI for a discussion.

## VI. DISCUSSION

### A. Vortex Propagation in Regular JTL, Circuit Revcom4

Using the design parameters in Table I we can estimate all parameters in (12) as given in Table II. Since $\Lambda_J/p > 2$, the JTL discreteness is small, there is no cell-to-cell energy barrier for a vortex motion, and the continuous junction approximation used to fit the CVCs is justified [37]. According to [39], [40], the single vortex flow resistance $R_0$ is related to the single junction damping resistance $R$ by

$$R_0 = \frac{\pi^2 R}{2N^2 \beta_L^{1/2}}, \tag{15}$$

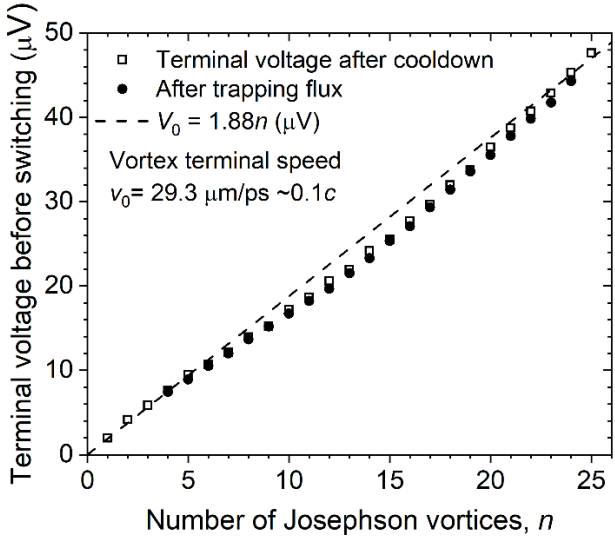

**Fig. 14.** Voltage on the nonuniform register Revcom5 right before switching from the vortex propagation state to the gap voltage of the ring at $I_{sw}$ at different number of Josephson vortices inserted into the register. Magnetic flux trapped in or around the circuit affects propagation of Josephson vortices, the shape of the CVCs, and the value of the terminal voltage.

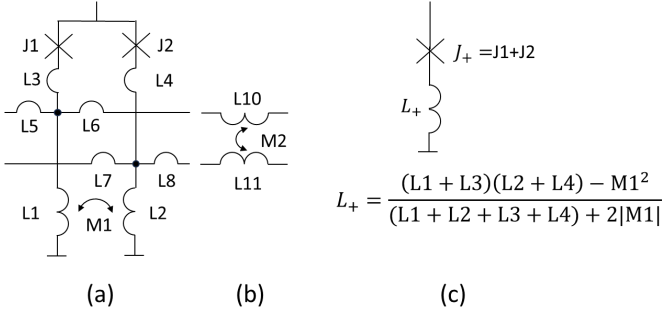

**Fig. 15.** (a) Equivalent circuits of nSQUID; (b) "negative inductance" nInd coupling the data lines; (c) a simplified single-JJ circuit representing the common mode of the nSQUID. Target parameters are J1=J2=8μA, L1=49.6 pH, L2= 48.9 pH, L3=1.6 pH, L4=2.3 pH, L5=L8=0.34 pH, L6=L7=1.1 pH, L10=L11=45 pH, M1=−38 pH, M2=−36.9 pH. They give $J_+$=16 μA and $L_+$=6.6 pH for the single-JJ circuit representing the common mode of the nSQUID in (c).

allowing us to estimate $R$ using $R_0$ in Fig. 9. This gives $R \approx 59.4$ kΩ, i.e., $R/R_N \approx 470$, which agrees with very high subgap resistances at low voltages observed in Nb-based JJs with similar $j_c$ fabricated at MIT LL for superconducting qubits [42], [43]. The corresponding $\beta_c \approx 1.5 \cdot 10^8$ indicates a highly underdamped system, allowing Josephson vortices propagate with extremely low energy dissipation, well below $E_T$ at propagation speeds $v \leq v_T \approx 0.7 v_0$.

### B. Vortex Propagation in nSQUID Discrete Transmission Line

Energy dissipation during information (Josephson vortices) movement in the nonuniform register composed of the regular and nSQUID JTLs is significantly higher than $E_T$ whereas it can be lower than $E_T$ in the regular JTL. Another striking difference is in depinning currents, likely indicating existence of a potential barrier for vortices between the regular JTL sections and the nSQUID sections, not anticipated in the design of the Revcom5 circuit. To explain these differences, we need to examine design of nSQUID-based discrete transmission lines.

A detailed circuit schematic of the nSQUID is shown in Fig. 15 along with all design parameters. For simplicity, the nSQUID operating in the common mode, without external flux coupled to loop L1-L3-J1-J2-L4-L2, can be replaced by a single JJ with critical current equal the sum of critical currents of the junctions J1and J2, $J_+ = J1 + J2 = 2I_c$ and inductance $L_+ = (L - |M|)/2$, where $L = (L1 + L3) = (L2 + L4)$ and $M$ is the mutual inductance value of M1 in Fig. 15.

Then, a discrete transmission line of nSQUIDs can be simplified as shown in Fig. 16. There are two substantial differences between the array of nSQUIDs and the regular JTL in Fig. 2: a) inductance $L_+$ above the JJs and the ground, which is comparable to the cell inductance $L_{cell}$ and, hence, cannot be neglected, contrary to a very small parasitic inductance $LJ_i$ in the regular JTL case; b) $C_{cell}$ is in parallel to the $J_+$-$L_+$ serial connection, not to the junction's $C_J$.

Kirchhoff's current equations for each node read

$$I_i = I_{i-1} + I_b - C_{cell} \frac{\partial V_i}{\partial t} - I_{Ji}, \tag{15}$$

$$I_{Ji} = I_+ \sin \varphi_i + \frac{\Phi_0}{2\pi R} \frac{\partial \varphi_i}{\partial t} + \frac{\Phi_0 C_J}{2\pi} \frac{\partial^2 \varphi_i}{\partial t^2}, \tag{16}$$

where

$$V_i = \frac{\Phi_0}{2\pi} \frac{\partial \varphi_i}{\partial t} + L_+ \frac{\partial I_{Ji}}{\partial t} \tag{17}$$

is voltage on the transmission line with respect to the ground; $I_{Ji}$ is current through the $i$-th nSQUID in the resistively capacitively shunted junction model, and $\varphi_i$ is the common mode phase difference across the $i$-th nSQUID. Currents and phases in (15), (16) are coupled by fluxoid quantization in each loop of the discrete transmission line

$$\varphi_{i-1} + \frac{2\pi L_+}{\Phi_0}\left(I_{Ji-1} - I_{Ji}\right) - \varphi_i - \frac{2\pi L_{cell}}{\Phi_0} I_{i-1} = 0 \tag{18}$$

---

### TABLE II
### EXTRACTED PARAMETERS OF VORTEX PROPAGATION

| Circuit | $R_0$ from CVCs (mΩ) | $R$, JJ damping resistance, (kΩ) | $\beta_c = 2\pi I_c R^2 C_J/\Phi_0$ | $c_0$, Swihart speed in different sections (μm/ps) | $V_0$ measured (μV) | $V_0$ calculated (μV) |
|---|---|---|---|---|---|---|
| Revcom4 | 10.7 | | 1.5E8 | 31.4[a] 26.2[b] | 4.2 | 5.0[a] 4.2[b] |
| Revcom5 Regular JTL section | | 59.4 | 1.5E8 | 46.0[b] | 1.88 | 4.0[a] 2.94[b] |
| Revcom5 nSQUID sections | | 59.4 | 1.5E8 | 46.0[b] 28.3[d] | 1.88 | 2.94[c] 1.81[d] 1.88[e] |

[a] Using junction specific capacitance of 40 fF/μm² from [45].
[b] Using junction specific capacitance of 59.3 fF/μm² that gives the best fit to the measured step voltage $V_0$ in the regular JTL ring Revcom4.
[c] Assuming a uniform propagation speed 46.0 μm/ps in all sections of the register.
[d] Assuming that $L_+$ adds in series to $L_{cell}$ and $C_{cell}$ is in parallel with $C_J$, and using the same propagation speed in the JTL and nSQUID sections.
[e] Assuming the maximum propagation speed of 28.3 μm/ps in nSQUID sections and a larger speed of 30.1 μm/ps in the JTL sections; see text.

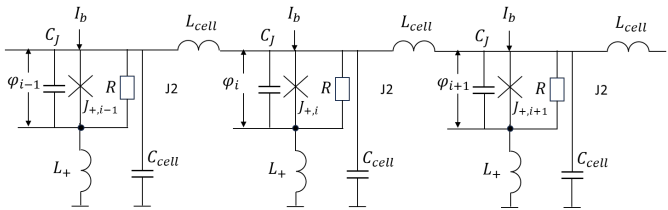

**Fig. 16.** A simplified circuit diagram of a few cells of the discrete transmission line composed of nSQUIDs operating in the common mode. Each nSQUID is replaced by a resistively capacitively shunted junction and a series inductance of the common mode $L_+$, as shown in Fig. 15, both in parallel to the cell capacitance to the ground.

$$\varphi_i + \frac{2\pi L_+}{\Phi_0}(I_{Ji} - I_{Ji+1}) - \varphi_{i+1} - \frac{2\pi L_{cell}}{\Phi_0}I_i = 0 \qquad (19)$$

and Kirchhoff's voltage equations

$$V_i - V_{i+1} - L_{cell}\frac{\partial I_i}{\partial t} = 0 \ . \qquad (20)$$

Equations (15)-(20) cannot be reduced to (12) and need to be solved numerically, e.g., using superconducting circuit simulators [46], [47] as will be presented elsewhere.

Some insight can be gained by considering a discrete passive $L$-$C$ transmission line consisting of series inductance $L_{cell}$ and a series connection of $C_J$ and $L_+$ to the ground. The cell capacitance and losses can be neglected since $C_{cell} \ll C_J$ and $R$ was found to be very large; see Tables I and II. Traveling electromagnetic waves $\exp i\,(\omega t \pm kx)$ on this transmission line have the propagation constant

$$|k| = \frac{\omega}{p}\left(\frac{L_{cell}C_J}{1-\omega^2 L_+ C_J}\right)^{1/2}, \qquad (21)$$

and the frequency-dependent propagation speed

$$v = \omega/k = v_0(1-\omega^2 L_+ C_J)^{1/2} \qquad (22)$$

which is always smaller than the propagation speed $v_0 = p/(L_{cell}\,C_J)^{1/2}$ on the regular transmission line with $L_+ = 0$. The wave impedance of this transmission line

$$Z = Z_0(1-\omega^2 L_+ C_J)^{1/2} \qquad (23)$$

is also frequency-dependent and always smaller than the wave impedance $Z_0 = \left(\frac{L_{cell}}{C_J}\right)^{1/2}$ of the regular transmission line with $L_+ = 0$.

This simplistic consideration suggests that the propagation speed in and impedance of the regular JTL and nSQUID JTL sections are different due to the presence of the series inductance to the ground. These differences may impede propagation of Josephson vortices on the mixed transmission line and increase energy dissipation, although the propagation cut-off frequency $1/2\pi(L_+ C_J)^{1/2} \approx 63$ GHz appears to be too high to cause a significant difference.

Empirically, the best agreement with the measured voltage step $V_0$ is obtained if $L_+$ is simply included in the total cell inductance of the nSQUID sections to set the maximum propagation speed in these sections as $v_0^{nSQUID} = p/[(L_{cell} + L_+)(C_J + C_{cell})]^{1/2} \approx 28.3$ μm/ps. A single vortex moving uniformly with this speed in the JTL and nSQUID sections would generate voltage $V_0 = 1.81$ μm, quite close to the measured value; see Table II. However, the vortex does not have to move with the constant speed. It can move faster in the JTL sections and slower in the nSQUID sections. The shape of the CVCs indicates that vortices quickly reach the maximum speed, apparently in the nSQUID sections, at currents slightly above the depinning current. Their speed in the JTL sections, $v^{JTL}$, can be estimated from the measured $V_0 = \frac{\Phi_0}{\tau} = 1.88$ μV, giving the vortex travel time around the ring $\tau$. The latter is related to the propagation speeds in the JTL and nSQUID sections, $\tau = \frac{l_{JTL}}{v^{JTL}} + l_{nSQUID}/v^{nSQUID}_0$, and the total lengths of the sections $l_{JTL}$ and $l_{nSQUID}$ given in Table I. This gives $v^{JTL} = 30.2$ μm/ps, a much lower value than the maximum possible, Swihart, speed in the JTL sections; see Table II.

So, it appears that vortices in the nonuniform register Revcom5 move nonuniformly, like cars on a variable-speed-limit highway, accelerating in the JTL sections and slowing down in the nSQUID sections, which causes much higher energy dissipation than in the uniform register Revcom4. Also, if adding $L_+$ to the cell inductance is justified in the future numerical simulations, this would increase the value of $\beta_L^{nSQUID}$ to 0.52 and decrease the size of the Josephson vortex $\Lambda_J^{nSQUID}$ to about 140 μm. At these values, discreteness of the nSQUID arrays becomes significant, which may cause a nonzero energy barrier between the adjacent cells and a finite vortex depinning current. These issues warrant further investigation.

## VII. Conclusion

We have studied propagation of information in the form of Josephson vortices in circular shift registers of two types: a) composed of sections of regular JTLs and b) composed of sections JTLs with nSQUIDs and sections of regular JTLs. We have found that energy dissipation in the uniform shift register can be lower than the Landauer's minimum energy $E_T$ required for computations at propagation speeds below about 0.7 of the maximum propagation speed, Swihart speed, on the Josephson transmission line. For the nonuniform register, the energy dissipation was much higher than in the uniform register and much higher than $E_T$, likely due to a mismatch of the impedances and propagation constants of the nSQUID-based sections and regular JTL sections of the register, and also to effects of flux trapping. We plan to perform a detailed numerical simulation of Josephson vortex propagation in the nonuniform registers in order to attempt to reproduce the observed features of the CVCs. The knowledge gained as a result of the presented measurements will be used to design nSQUID-based logic cells for reversible computing.

## Acknowledgment

The circuits studied in this work were fabricated in the Microelectronics Lab of the MIT Lincoln Laboratory using multiproject fabrication runs of the SFQ5ee process. We are grateful to Vladimir Bolkhovsky, Ravi Rastogi, and David Kim for overseeing these fabrication runs.

Any opinions, findings, conclusions, or recommendations